\documentclass[twocolumn,showpacs,preprintnumbers,amsmath,amssymb]{revtex4}
\def\beq{\begin{equation}}
\def\eeq{\end{equation}}
\def\beqa{\begin{eqnarray}}
\def\eeqa{\end{eqnarray}}

\def\l{\left}
\def\rr{\right}
\def\bdi{\begin{displaymath}}
\def\edi{\end{displaymath}}

\usepackage{graphicx}
\usepackage{dcolumn}
\usepackage{bm}
\begin{document}

\preprint{APS/123-QED}
 
\title{Elasticity of semiflexible polymers with and without self-interactions}
\author{A. Rosa$^{1}$, T. X. Hoang$^{2}$, D. Marenduzzo$^{3}$, A. Maritan$^{1,2}$} 
\affiliation{
$^1$ International School for Advanced Studies (SISSA) and INFM, Via Beirut 2-4, 34014 Trieste, Italy\\ 
$^2$ The Abdus Salam International Center for Theoretical Physics (ICTP), Strada Costiera 11, 34100 Trieste, Italy\\
$^3$ Department of Physics, Theoretical Physics, University of Oxford, 1 Keble Road, Oxford OX1 3NP, England\\ 
}

\begin{abstract}
A {\it new} formula for the force vs extension relation is derived
from the discrete version of the so called {\it worm like chain}
model. This formula correctly fits some recent experimental data on
polymer stretching and some numerical simulations with pairwise
repulsive potentials. For a more realistic Lennard-Jones potential the agreement with simulations is found to be good when the
temperature is above  the $\theta$ temperature. For lower
temperatures a plateau emerges, as predicted by  some recent
experimental and theoretical results, and our formula gives good 
results only in the high force regime. We briefly discuss how
other kinds of self-interactions are expected to affect the elasticity of the polymer. 
\end{abstract}


\maketitle

\begin{section}{Introduction}\label{intro}
The behaviour of a single polymeric molecule under stretching has become a 
popular subject among experimental physicists (see e.g. \cite{bustamante}, 
for a review) and has attracted the attention of theoreticians who have 
introduced several models to explain these results. 
An exhaustive description on this topic is hard to achieve within a single
model, since the experimental conditions strongly affect the results. So,
the most common theoretical approaches are usually only a first step approximation.    

Here, we focus our attention mainly on the pulling of a polymer under good 
solvent conditions \cite{degennes}, in which case we give an analytical
prediction for the force versus extension curve. 
We also discuss other cases by means of 
numerical simulations and direct connections to experiments. Usually, the 
models adopted describe the polymer as an elastic (ideal) chain, where 
self-avoidance is not taken into account. Typically, freely jointed (FJC) or 
worm like chain (WLC) are studied \cite{wang}. 

The former describes a chain of beads connected by links of constant distance, whereas the latter introduces an intrinsic {\it stiffness} between two consecutive bonds and, in particular, is shown to correctly
describe a wide range of experimental  results on double-stranded
(ds) DNA, single plasmid and lambda phage DNA \cite{markosiggia,
baumann,podg1}. In this case, the large force behaviour  $1-z/L_c \sim
1/\sqrt{F}$ is found, where $z$ is the elongation along the 
direction of the force, $L_c$ the contour length of the polymer and
$F$ is  the applied force \cite{markosiggia,lamura}. Let us note that in
these cases  the {\it continuous} version of the WLC model is always
used, where the {\it  persistence} length \cite{markosiggia} $L_p$ is
very large, compared to the  base separation (roughly one
persistence length is 150 base pairs). 

Here, we focus on the {\it discrete} version of the WLC model, that 
satisfactorily describes the pulling behaviour of a polymer in good 
solvent. 
We broadly identify three regimes in the force vs
extension curves obtained in our analytical and numerical calculations.
The low force (or low stretch) regime is highly affected by the details
of the interactions between the beads (which in nature are, e.g., caused by the different concentration of ions
in solution). This regime is discussed only marginally here, as, though it is potentially very interesting, experimental data in this
range of forces are quite rare and not precise enough to allow a comparison with
theoretical predictions. There is then a second regime, of intermediate
stretches or forces, in which the force versus extension
characteristic curves obey approximately the laws predicted 
a few years ago in Ref. \cite{markosiggia} by means of a 
continuum theory of the WLC. Finally, for very large forces (beyond a polymer
dependent crossover value), we get a universal, model-independent,
freely-jointed-chain like behaviour.
In Section \ref{model}, we introduce the discrete WLC model and show
how it can  be related to the well known continuous version.
In Section \ref{experiments}, we show a comparison between our theory and some recent 
experiments. 
Then, in Sections \ref{mcarlo} and \ref{mdyn}, we introduce 
some extension vs force curves obtained from Monte-Carlo and Molecular 
Dynamics calculations for models of stiff and flexible polymers with a chosen potential 
between {\it non} nearest neighbour beads. We shall consider two different 
cases: a pure repulsive potential and a more realistic
Lennard-Jones potential  and we show that in the latter case our
formula agrees with the numerical  data only in the case of high
temperatures or, for lower temperatures in the  high force regime.
Below the $\theta$ point a force plateau appears, as  predicted by
some recent theoretical works \cite{halperin,prl,murayama,vilgis} and 
confirmed by experiments \cite{baumann}. Finally, Sections \ref{discussion} and \ref{conclusions} are
left for discussion and conclusions, respectively.

\end{section}

\begin{section}{The model}\label{model}
Our model describes a chain of beads, where the distance between the nearest 
neighbours is kept fixed (we can put it equal to {\bf $b$}) and a
suitable  {\it stiffness} $K$ is introduced. Then, the corresponding
Boltzmann weight reads \cite{fixman,coccorna1,coccorna2}:
\beq\label{wlc_model}
e^{-\beta\mathcal{H}} = \prod_{i=1}^N\delta(|{\mathbf t}_i|-b)
e^{\beta K\sum_{i=1}^N  {\mathbf t}_i {\mathbf t}_{i+1} +
\beta{\mathbf F} \sum_{i=1}^N {\mathbf t}_i}, 
\eeq 
where $\beta = 1/T$, $T$ being the temperature in units of
Boltzmann constant,  ${\mathbf t}_i \equiv {\mathbf r}_{i}-{\mathbf
r}_{i-1}$ (${\mathbf r}_i$ being the  position vector for the $i$-th
bead, {\bf  $i = 1, ..., N$} and $N$ is the total number of  beads)
and {$\mathbf F$} is the applied force which defines the 
$z$-direction. The partition function for the model described by
(\ref{wlc_model}) is
\beq\label{partfunc} \mathcal{Z}_N =
\int{\prod_{i=1}^N}d{\mathbf t}_i e^{-\beta\mathcal{H}}. 
\eeq 
The average elongation $\langle z \rangle_N$ along the stretching
direction is 
\beq\label{elong} 
\langle z \rangle_N = T \frac{\partial}{\partial F}\log\mathcal{Z}_N.   
\eeq  
In the thermodynamic limit ($N \rightarrow \infty$), the large force behaviour
of  Eq. (\ref{elong}) gives 
\beq\label{largeforce1} 
\zeta \equiv \lim_{N\rightarrow \infty} \frac{\langle z \rangle_N}{N b} =
1- \frac{T}{\sqrt{(b F)^2 + 4 b K F}}. 
\eeq  
The continuum approximation of Eq. (\ref{largeforce1}) is obtained with the
following substitutions \cite{dohmi}: 
\beq\label{cont_approx}
\beta K \rightarrow L_p/b,
\eeq
in the formal limit $b \rightarrow 0$, where $L_p$ is the persistence
length. The final result $\zeta = 1 - 1/2\sqrt{\beta L_p F}$ does agree 
with the celebrated result of Marko and Siggia \cite{markosiggia}. However, our
result is more general, since it predicts a {\it crossover} force $F_c = \frac{4
L_p}{\beta b^2}$:  
\beq\label{crossover}
\begin{array}{ccc}
\mbox{WLC-like behaviour:} & 1-\zeta \sim 1/\sqrt{F}, & F \ll F_c\\
{} & {} & {}\\
\mbox{FJC-like behaviour:} & 1-\zeta \sim 1/F, & F \gg F_c
\end{array}
.
\eeq
Let us notice that the validity of the continuum approximation proposed in Ref.
\cite{markosiggia} is not simply related to the value of the dimensionless
ratio $b/L_p$ but rather to $F/F_c$.

From Eq. (\ref{largeforce1}), we
deduce that:  
\beq\label{largeforce2}
\beta b F = \frac{2 L_p}{b}\l[\sqrt{1+\l(\frac{b}{2L_p}\rr)^2\frac{1}
{(1-\zeta)^2}}-1\rr].
\eeq
The low force behaviour of Eq. (\ref{elong}) is 
\beq\label{lowforce}
\zeta = \frac{\beta b F}{3}\frac{1+y(L_p/b)}{1-y(L_p/b)},
\eeq
where $y(x) = \coth(x)-1/x$ \cite{doi}. Again, in the limit $b \rightarrow 0$, 
Eq. (\ref{lowforce}) agrees with the result of Marko and Siggia \cite{markosiggia}. 
Matching Eqs. (\ref{largeforce2}) and (\ref{lowforce}), we obtain the following 
interpolation formula:
\beqa\label{interpol}
\beta b F & = & \frac{2 L_p}{b}\l[\sqrt{1+\l(\frac{b}{2L_p}\rr)^2\frac{1}
{(1-\zeta)^2}}-\sqrt{1+\l(\frac{b}{2L_p}\rr)^2}\rr] \nonumber\\
& & +\l(3\frac{1-y(L_p/b)}{1+y(L_p/b)}-\frac{b/2L_p}{\sqrt{1+(b/2L_p)^2}}\rr)\zeta.
\eeqa
It is easy to verify that Eq. (\ref{interpol}) correctly reproduces the right
large and small force behaviours, Eqs. (\ref{largeforce2}) and
(\ref{lowforce}) and that in  the {\it continuum} limit $b \rightarrow 0$ we
obtain the well known interpolation formula \cite{markosiggia}:
\beq\label{markosiggia} \beta L_p F = \frac{1}{4(1-\zeta)^2} - \frac{1}{4} +
\zeta. 
\eeq

In next Section, we shall apply our formula, Eq. (\ref{interpol}), to two recent 
experiments. The first discusses the stretching of a single plasmid DNA molecules, to which the formula
(\ref{markosiggia}) was previously applied with success \cite{baumann}. 
Nevertheless, our formula predicts a non trivial value for the parameter $b$, 
that gives an estimate for the {\it intra} bead distance. 
Then, the second experiment \cite{rief,dessinges} demonstrates that Eq. (\ref{interpol}) gives the right large force behaviour.

In the following, we 
shall define $\zeta = z/L_c$, where $z$ is the elongation along the direction of 
the force and {\bf $L_c = N b$} is the contour length of the polymer (see also
Eq. (\ref{largeforce1})).   
\end{section}

\begin{section}{Comparison with two recent experiments}\label{experiments}

Let us consider the experimental data reported in 
the plot at the top of Fig. 3 of Ref. \cite{baumann}. The authors 
considered the elastic response of a single plasmid of DNA molecules, probed 
using optical tweezers \cite{bockelmann}. They found that, according to the 
environmental conditions, it can be very different. In particular, for 
condensed molecules, the stretching curves display a stick-release pattern, 
where the DNA molecule can be described as a succession of different WLC's 
of different contour and persistence lengths. We are mainly interested to
the case of uncondensed molecules, whose stretching pattern is reproduced in
Fig. \ref{baumann_fig}. 
\begin{figure}
\includegraphics[width=2.5in]{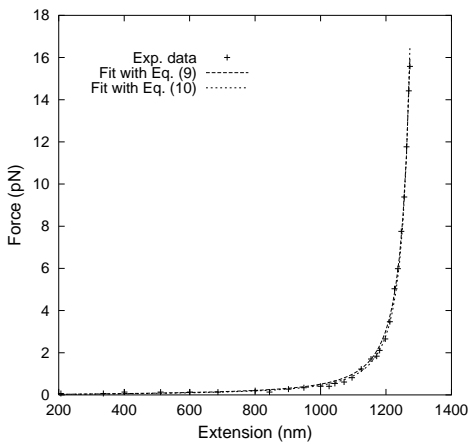}
\caption{(+): experimental data taken from Ref. \cite{baumann}. 
Long dashed line: fit with the curve, Eq. (\ref{interpol}), yielding $L_c=1324$ nm, 
$L_p=38$ nm and $b=2.5$ nm. 
Short dashed line: fit with the curve, Eq. (\ref{markosiggia}), yielding $L_c=1324$ 
nm and $L_p=38$ nm, see Ref. \cite{baumann}.}
\label{baumann_fig}
\end{figure}

As described in the caption, the contour and persistence lengths obtained 
with our formula, Eq. (\ref{interpol}), and with Eq. (\ref{markosiggia}) 
are perfectly compatible. Nevertheless Eq. (\ref{interpol}) predicts a non 
trivial value for the intra bead effective distance $b=2.5$ nm, that corresponds 
to $7-8$ base pairs. Let us notice that this matches the
DNA hydration thickness (here, we have
used $\sim 0.34$ nm as the distance between two consecutive base pairs \cite{baumann}).

If we take $T = 300$ K, the crossover force $F_c \sim 100$ pN and 
$F/F_c \lesssim 0.16$ for the data of Ref. \cite{baumann}, thus justifying the
use of the interpolation formula, Eq. (\ref{markosiggia}).
However, as pointed out in Eq. (\ref{crossover}), the discrete nature of the
chain does emerge, when $F \gtrsim  F_c$. Notice also that for forces
considerably smaller than $F_c$ double-stranded DNA would undergo an
overstretching transition \cite{overstretch}, where a more sophisticated theory is needed \cite{wang}.

An interesting question is how this treatment may be applied to
single-stranded DNA (ssDNA). On one hand, if we keep as physical parameters
the persistence length of ssDNA $\sim 1$ nm, and as the equivalent of $b$ the
separation between two phosphates, i.e. $0.5$ nm roughly, we would end up
with a crossover force again of the order of 70 pN. Data in this regime 
do exist \cite{rief,dessinges}, and suggest that the WLC 
grossly fails to fit the data \cite{dessinges,mezard}. In fact (see Fig. 4, Ref. \cite{dessinges}), the authors pointed out that the corresponding fit with Eq. \ref{markosiggia} gives good results in the large force regime, but with a calculated persistence length ($\simeq 0.21$ nm) which is clearly {\it not} physical. Our equation does not
do much better for low and intermediate force, in which case, as 
shown in Ref. \cite{dessinges}, evidently the self-interactions dominate the
behaviour. Still a large force fit, even if a bit dependent 
on the contour length which we choose, suggests that the large
force exponent in the $(\log(1-\zeta),\log(F))$ plane is $-1$ as predicted by our model (see Fig. \ref{extra}). 
\begin{figure}
\includegraphics[width=2.5in]{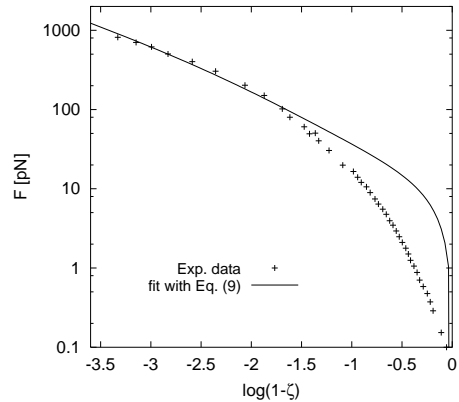}
\caption{(+): experimental data taken from Ref. \cite{dessinges}. 
Continuous line: fit with the curve, Eq. (\ref{interpol}), yielding $L_c=2.31$ (in units of the contour length $l^0_{ds}$ of the equivalent dsDNA molecule observed in $10$ mM PB \cite{dessinges}), $L_p=0.26$ nm and $b=0.12$ nm.}
\label{extra}
\end{figure}
The calculated fitting parameters $L_p$ and $b$ (see the caption of Fig. \ref{extra}) give $F_c \simeq 300$ pN, which is consistent with our Eq. (\ref{interpol}) (see Fig. \ref{extra}).

In next Section we introduce some Monte-Carlo calculations and compare them to Eqs. (\ref{interpol}) and (\ref{markosiggia}).

\end{section}

\begin{section}{Monte-Carlo calculations}\label{mcarlo}
As already said, our model is a
{\it stiff} chain described by the Boltzmann weight,  Eq. (\ref{wlc_model}),
where the {\it intra} bead distance $b$ is now kept fixed to $1$.

In Fig. \ref{pure_wlc} we have plotted the Monte-Carlo data (+) and the curves 
given by Eqs. (\ref{interpol}) and (\ref{markosiggia}) (long and short dashed line, 
respectively), for $L_p=40$, and $\beta=1$. As 
observed, the agreement is perfect only for Eq. (\ref{interpol}). In fact, the 
discrete nature of the chain emerges around $F = F_c \sim 160$ (in the
chosen units) and the WLC approximation is no more valid.
\begin{figure}
\includegraphics[width=2.5in]{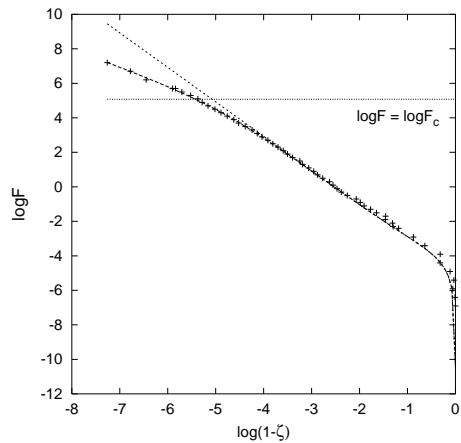}
\caption{(+): Monte-Carlo data for a {\it stiff} chain of $N=100$ beads and 
stiffness $K=40$, for $\beta=1$. Long dashed line: Eq. (\ref{interpol}) with 
$L_p = 40$ and $b=1$. Short dashed line: Eq. (\ref{markosiggia}) with $L_p = 40$.
 The {\it crossover} force $F_c$, see Eq. (\ref{crossover}), is also shown.}
\label{pure_wlc}
\end{figure}

To render the model more realistic, 
let us introduce a {\it short} range repulsive potential 
$V_{rep}(r)$ between non consecutive nearest neighbour beads. We have adopted the 
following functional form:
\beq\label{repulsive}
V_{rep}(r) = \frac{1}{r^{12}}
\eeq
where $r$ is the distance between two beads.
Then, we have to multiply the Boltzmann weight, Eq. (\ref{wlc_model}), by 
\beq\label{rep_weight}
e^{-\frac{\beta}{2}\sum_{i \neq j=0}^N V_{rep}(r_{ij})} 
\eeq 
where we have defined $r_{ij} = |{\mathbf r}_i - {\mathbf r}_j|$. 

In Fig. \ref{rep_wlc} we have plotted the Monte-Carlo results (+) for $K=30$, 
together with the two fitting lines obtained from Eqs. (\ref{interpol}) and 
(\ref{markosiggia}) (long and short dashed lines, respectively). The corresponding 
persistence lengths are $L_p \simeq 31$ ($b$ is kept fixed to 1) and 
$L_p \simeq 34.9$. Again, we can observe that our formula works better than Eq. 
(\ref{markosiggia}). Moreover, this last result is in agreement with some 
theoretical works \cite{thirumalai,podg}, that predicts that the net
effect of the repulsive potential is to renormalize the persistence length, making it larger than the bare one.

\begin{figure} \includegraphics[width=2.5in]{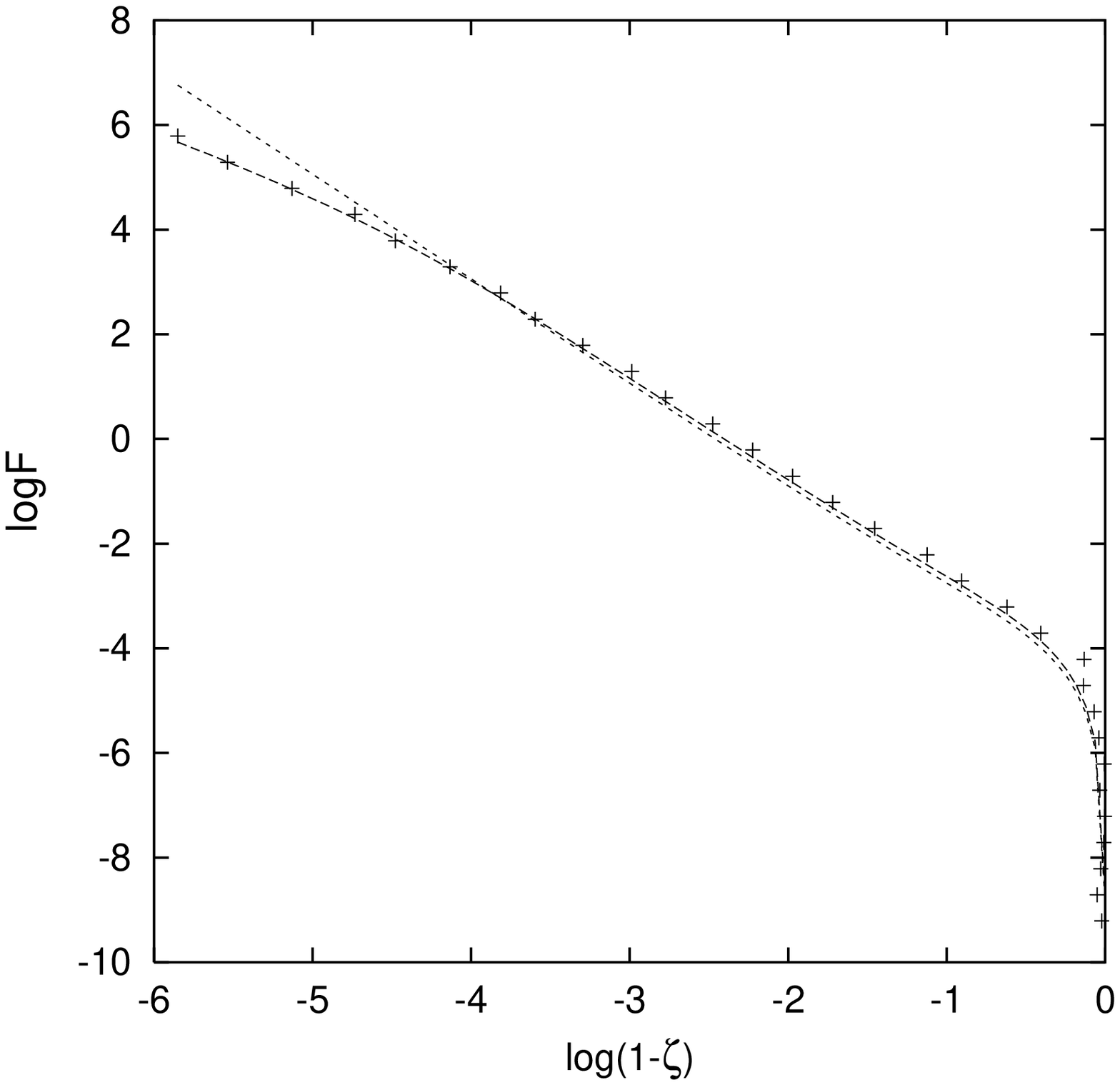}
\caption{(+): Monte-Carlo data for a {\it stiff} chain of $N=100$ beads, stiffness 
$K=30$ and interaction potential given by Eq. (\ref{repulsive}), for $\beta=1$. 
Long dashed line: fit obtained from Eq. (\ref{interpol}) with $L_p \simeq 31$ 
($b=1$). Short dashed line: fit obtained from Eq. (\ref{markosiggia}) with 
$L_p \simeq 34.9$.}
\label{rep_wlc}
\end{figure}

Now, let us introduce a more realistic potential, adding to the repulsive core an 
attractive part too. It is interesting to notice that the approaches of Refs. \cite{thirumalai,podg} can not be generalized, since the attractive part introduces some instabilities and the perturbation scheme discussed there is no more valid. 

We have chosen the following Lennard-Jones kind
functional form  $V_{LJ}(r)$: \beq\label{lennjones} V_{LJ}(r) =
V_0\l(\frac{1}{r^{12}} - \frac{\alpha}{r^6}\rr). \eeq The parameters $V_0$ and
$\alpha$ are chosen in such a way that the minimum of the  energy is located at
$r=r_{min}=1.5$ and $V_{LJ}(r=r_{min})=-1$. We expect that  Eq.
(\ref{markosiggia}) does not work well in this case. Our goal is to test our 
formula, Eq. (\ref{interpol}).

Firstly, we show the effects of increasing stiffness on the force vs extension 
curves. To fix the ideas, we begin to consider a sufficiently high temperature, 
i.e. {\it above} the $\theta$ point \cite{degennes,doi}, whose location, however, 
is not exactly known. In Ref. \cite{karplus}, the authors were able to numerically 
determine the $\theta$ temperature $T_{\theta}$ for a model of homopolymer with a 
square well potential, whose depth is fixed to $-1$. They found that $T_{\theta} 
\simeq 3$. Our $T_{\theta}$ should be somewhat smaller due to the stiffness.

Initially, we fix $\beta=0.3$ ($T=3.333...$), with $K=10$. The numerical results and the corresponding fitting
line are plotted in Fig. \ref{beta03.1}. 
\begin{figure}
\includegraphics[width=2.5in]{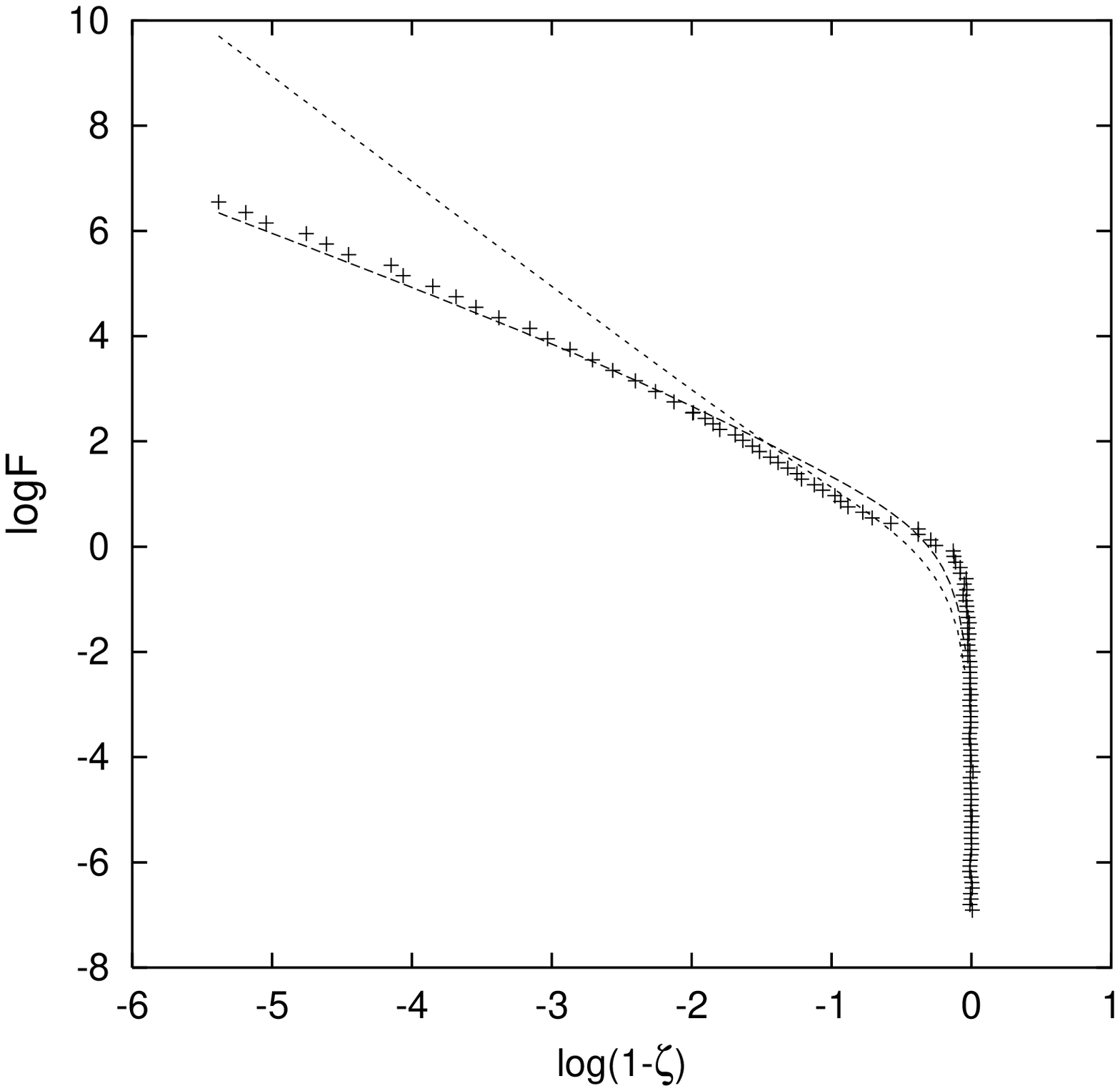}
\caption{(+): Monte-Carlo data for a {\it stiff} chain of $N=100$ beads, stiffness 
$K=10$ and interaction potential given by Eq. (\ref{lennjones}), for $\beta=0.3$. 
Long dashed line: fit obtained from Eq. (\ref{interpol}) with $L_p \simeq 2.12$, 
$b \simeq 1.26$. Short dashed line: fit obtained from Eq. (\ref{markosiggia}) 
with $L_p \simeq 2.41$.}
\label{beta03.1}
\end{figure}
The corresponding fitting parameters are described in the caption. 
In this case we allowed $b$ to be a free parameter and we have determined it
through a best fit of the data even though in the simulated model $b = 1$.
The fit with our formula is surprisingly good, in contrast with
the result obtained  with Eq. (\ref{markosiggia}). Moreover, the predicted value for $b$ is of the correct order. 
This means that the effect of the
stiffness compensates the attraction due to the potential and the behaviour is
similar to a FJC. Let us stress on the fact that if we fix $b=1$, the corresponding fits are considerably less precise.

We have also simulated the case with $K=80$. In Fig. \ref{beta03.2} we
have plotted the numerical data  together with the two fitting lines (the
fitting parameters are reported in the caption). 
\begin{figure}
\includegraphics[width=2.5in]{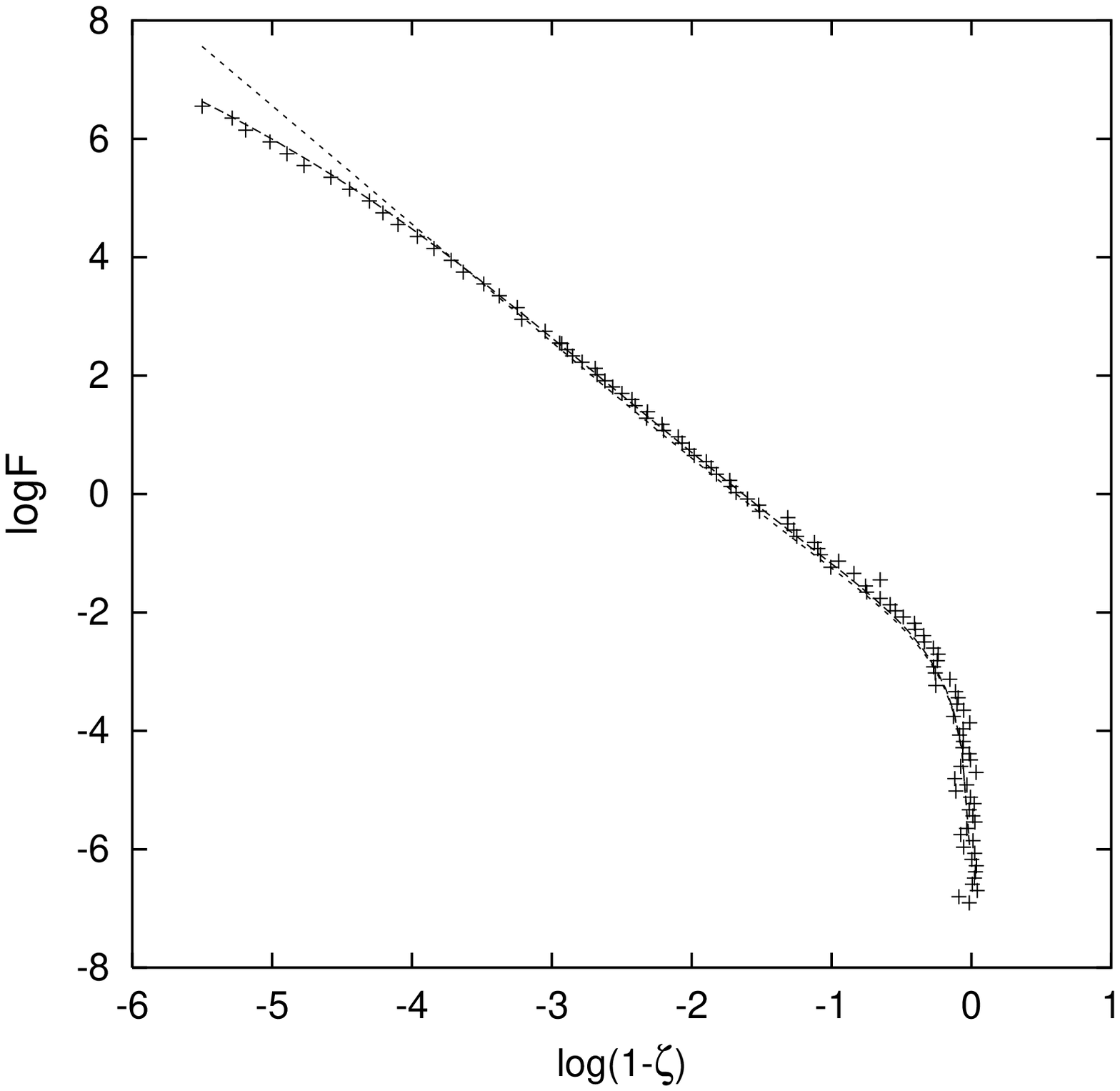}
\caption{(+): Monte-Carlo data for a {\it stiff} chain of $N=100$ beads, stiffness 
$K=80$ and interaction potential given by Eq. (\ref{lennjones}), for $\beta=0.3$. 
Long dashed line: fit obtained from Eq. (\ref{interpol}) with $L_p \simeq 23.33$ 
and $b=0.86$. Short dashed line: fit obtained from Eq. (\ref{markosiggia}) with 
$L_p \simeq 25.95$.}
\label{beta03.2}
\end{figure}
The agreement with our formula is again perfect, in contrast with Eq. (\ref{markosiggia}). 

For both situations plotted in Figs. \ref{beta03.1} and \ref{beta03.2}, we note 
that the main effect due to the potential is a considerable reduction in the 
persistence length due to the attractive part, which evidently renormalize also 
the intra bead distance.

As a check, let us notice that the correlation function $\langle {\mathbf t}_i \cdot {\mathbf t}_{i+r} \rangle$ for the WLC is \cite{doi}:
\beq\label{corrfunct}
\langle {\mathbf t}_i \cdot {\mathbf t}_{i+r} \rangle = y(\beta K)^r,
\eeq
where the function $y(x)$ has been defined above. In the continuum limit 
\beq\label{contcorrfunct}
\langle {\mathbf t}_i \cdot {\mathbf t}_{i+r} \rangle = \exp(-r/L_p). 
\eeq
This result is exact for all $r$ but holds in a more general context, like the case with interaction, in the large $r$ limit. In Fig. \ref{WLCcorr} we have plotted the correlation functions $\langle {\mathbf t}_1 \cdot {\mathbf t}_{1+r} \rangle$ vs $r$ for the data of Figs. \ref{pure_wlc} to \ref{beta03.2} and compared them to the theoretical ones as given by Eq. (\ref{contcorrfunct}). The corresponding persistence lengths are those found with our formula. The agreement is perfect. 

\begin{figure}
\includegraphics[width=2.5in]{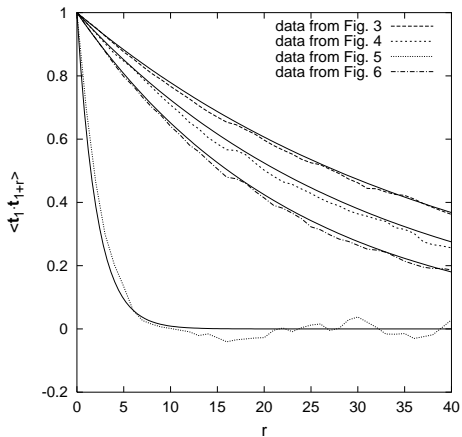}
\caption{Plot of the correlation functions $\langle {\mathbf t}_1 \cdot {\mathbf t}_{1+r} \rangle$ vs $r$, calculated from the data of Figs. \ref{pure_wlc} to \ref{beta03.2} and the corresponding theoretical predictions (continuous lines), given by Eq. (\ref{contcorrfunct}).}
\label{WLCcorr}
\end{figure}

Let us now consider the case of $\beta = 0.5$ ($T = 2$). In Fig. \ref{beta05} we 
have plotted the case for $K=10$.
\begin{figure}
\includegraphics[width=2.5in]{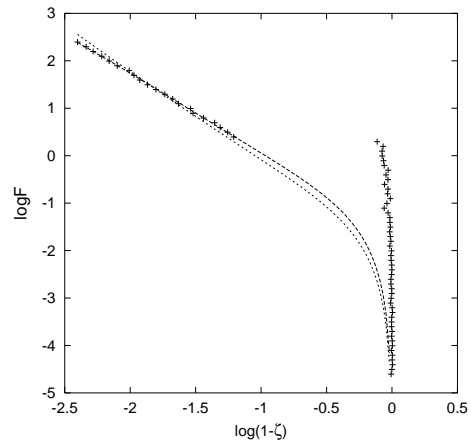}
\caption{(+): Monte-Carlo data for a {\it stiff} chain of $N=100$ beads, stiffness 
$K=10$ and interaction potential given by Eq. (\ref{lennjones}), for $\beta=0.5$. 
Long dashed line: fit obtained from Eq. (\ref{interpol}) with $L_p \simeq 4.42$ 
and $b = 1.03$. Short dashed line: fit obtained from Eq. (\ref{markosiggia}) with 
$L_p \simeq 4.85$. Let us remark that we have tried to fit only the range of data 
at large forces.}
\label{beta05}
\end{figure}
As it can be seen a {\it first order} phase transition emerges
\cite{prl,pre}. In this  case our formula is able to describe only the
part of the plot at relatively large forces.

A complete characterization of the force vs extension behaviour in this case needs 
a more complete theory and is beyond the scopes of this work. We reserve to study 
this important issue in a future work (see also Ref. \cite{kessler} for some recent Monte-Carlo results about stiff polymers).
\end{section}

\begin{section}{Molecular Dynamics calculations}\label{mdyn}

In the Monte-Carlo simulations presented in
previous Section we were working in fixed force ensembles,
i.e. the force was fixed in each simulation and the averaged
extension of the chain was computed. It is interesting to check if
in fixed stretch ensembles, i.e. when one fixes the extension and
measures the averaged force, the results remain unaltered. In the
following we will present and discuss Molecular Dynamics (MD) simulation
results in which we can perform a quasi-fixed stretch treatment.
In these simulations (similar to what has been done for
protein-like polymers \cite{Hoang1,Hoang2}) 
the two ends of the polymer are tethered 
by springs to two fixed points in space so its end-to-end vector
can vary only little. The two fixed points are chosen such that
a line connecting them is along the $z$-direction. 
We will consider two cases: a) when
there are repulsions between two non-consecutive beads and b)
when the beads interact via a Lennard-Jones (LJ) potential.
Note that we do not include the 
intrinsic stiffness into the models considered in this Section (i.e. $K=0$).

Our homopolymers are modeled as chains of beads connected by
springs. The harmonic potential for the chain connectivity reads
\begin{equation}
V_{i,i+1} = \frac{1}{2}k(r_{i,i+1}-b)^2 \;,
\end{equation}
where $r_{i,i+1}=|{\bf r}_i - {\bf r}_{i+1}|$ is the distance
between two consecutive beads. $b$ is fixed to $1$
and $k=1444\epsilon$ is the chosen spring constant
($\epsilon$ is a unit of energy). 
We use the same spring constant $k$ for the potential
that tethers the ending beads to the fixed points. 
For the case of repulsion the following potential is used
\begin{equation}
V_{i,j} = \epsilon \left(\frac{\sigma}{r_{ij}}\right)^{12}\;\;,
\label{rp_pot}
\end{equation}
where $\sigma = 1.32$. For the case of attraction the LJ potential
takes the form
\begin{equation}
V_{i,j} = 4\epsilon\left[\left(\frac{\sigma}{r_{ij}}\right)^{12}
-\left(\frac{\sigma}{r_{ij}}\right)^{6}\right] \;.
\label{lj_pot}
\end{equation}
The model is studied by a MD method in which coupling to the heat
bath is provided by the Langevin equation (see e.g. Ref. \cite{Hoang1} for
details). In the following temperature will be given in units
of $\epsilon/k_B$ where $k_B$ is the Boltzmann constant.
We consider chains of 60 residues and simulations are done
for various temperatures. 
The averaged force $\left<F\right>$ exerted on each of the 
fixed points and the averaged $z$-component of the 
end-to-end vector of the chain $\left<z\right>$
are measured under equilibrium condition at a constant temperature.
As we vary the position of
one fixed point versus the other we are able to obtain $\left<F\right>$
vs $\left<z\right>$ under a quasi-fixed stretch condition (the 
fluctuation in $z$ is considerably small).

It should be noted that due to the springs used for the chain connectivity
the contour length $L_c$ is not a constant 
as one varies the stretching distance. We find that as the
polymer approaches its full extension $L_c$ may
increase up to 5-10\% depending on the temperature.
We find that however if
the relative extension $\zeta$ is defined as 
$\left<z\right>$/$\left<L_c\right>$, where $\left<L_c\right>$
is the averaged contour length computed from the simulations 
at a given stretch $z$, then the $\left<F\right>$ vs $\zeta$ 
acquires very little changes as we increase $k$ by a factor of 2 or 3.

\begin{figure}
\includegraphics[width=2.5in]{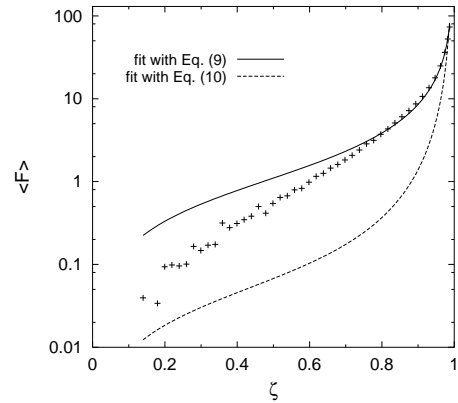}
\caption{Force $\left<F\right>$ vs relative extension 
$\zeta$ for a 60-bead chain with
repulsion studied by Molecular Dynamics simulations under
quasi-fixed stretch conditions (see text). 
$\zeta$ is defined as $\langle z \rangle$/$\langle L_c \rangle$
where $\langle z \rangle$ and $\langle L_c \rangle$ are the averaged 
extension and the averaged contour length obtained in the
simulation.  The data points shown are for temperature $T=1$.
The fit to Eq. (\ref{interpol}) (solid line) yields
$b\simeq 1.05$ and $L_p \simeq 1.04$. Eq. (\ref{markosiggia}) fails to fit
the data (dashed line).}
\label{prp60}
\end{figure}

In Fig. \ref{prp60} we show the stretching data for the chain with repulsive
potential given by Eq. (\ref{rp_pot}) at $T=1$ (at some other temperatures
we found that the results are qualitatively similar). One
can see that the data can be fitted quite well by Eq. (\ref{interpol})
at high extension, but not at low extension. 
The least-squared fit to the data gives $b \simeq 1.05$ and 
$L_p \simeq 1.04$.  
Note that the value of $b$ obtained from the fit is reasonably good
though it is somewhat larger than
the bond length when no tension is applied.
This increment (of about 5\%) however is in a good agreement with
the increase in $L_c$ under stretching and can be understood as due to the 
softness of the bond's potential. Thus Eq. (\ref{interpol}) still gives 
the right physics.
One can also notice that our homopolymer with repulsive potentials
corresponds to a worm-like chain with a low stiffness ($L_p \simeq b$).
We have computed directly the persistence
length from the correlation $\langle{\bf t}_i \cdot {\bf t}_{i+r}\rangle$ 
by performing simulations for a free chain
and found that $L_p$ is approximately $2.35$ at $T=1$. 
The difference (of a factor of $2$) between the measured persistence length
and the one obtained from the fit is due to the fact that the fit
was good only at a high extension. In this regime the excluded
volume effects are much smaller than in a free chain, thus
resulting in a lower persistence length.
It should be noted that Marko and Siggia's formula (Eq. (\ref{markosiggia})) 
completely fails to fit the data.
This is related to the fact that for chains with low stiffness 
the continuum limit $b \rightarrow 0$ cannot be applied and the 
discreteness of the chain should be taken into account.

\begin{figure}
\includegraphics[width=2.5in]{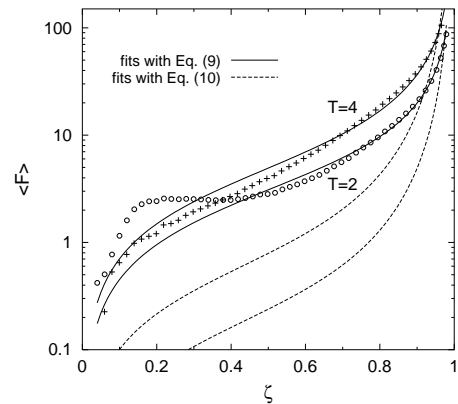}
\caption{Force $\langle F \rangle$ vs relative extension $\zeta$ for 
chain with Lennard-Jones (LJ) potentials under quasi-fixed stretch
conditions. The fits of the data to Eq. (\ref{interpol}) (solid lines) 
yield $b \simeq 1.01$ and $L_p \simeq 0.88$ for $T=4$, and
$b \simeq 1.0$ and $L_p \simeq 0.5$ for $T=2$.
Eq. (\ref{markosiggia}) fails to fit the data at
both temperatures (dashed lines).}
\label{plj60}
\end{figure}

We turn now to discuss our results for homopolymers with attraction.
Fig. \ref{plj60} shows the force vs relative extension for 
a 60-bead chain with the Lennard-Jones potential given by Eq. (\ref{lj_pot}).
We find that Eq. (\ref{interpol}) can reasonably fit the data for 
temperatures higher than the collapse transition temperature $T_\theta$ 
whereas at lower temperatures it can fit the data only at high extension.
The $T_\theta$ temperature is estimated to be between 2 and 3 
for this system. As shown in Fig. \ref{plj60} 
at $T=4$ Eq. (\ref{interpol}) appears to fit both high and low extension data 
yielding $b \simeq 1.01$ and $L_p \simeq 0.88$. At $T=2$ the
fit can be given only at large extension, which yields $b \simeq 1.0$ 
and $L_p \simeq 0.5$. Note that at both temperatures the value of $b$ 
obtained from the fits is quite consistent with the correct bond length.
Like for the case of repulsion, due to the low
stiffness induced by the potentials, the Marko and Siggia formula does 
not fit the data as our new formula does. 

It should be noted that when the temperature is about the collapse transition
temperature $T_\theta$ we observe a {\it plateau} in the force vs extension
(see the case of $T=2$ in Fig. \ref{plj60}), similar to what seen in the previous Section. 
However, at temperatures lower than $T_\theta$, in the fixed stretch ensemble the force vs extension curve shows the typical behaviour with hysteresis phenomenon \cite{Hoang1,Hoang2}.

\end{section}

\begin{section}{Discussion}\label{discussion}
Though our main results refer to a situation in which attractive
interactions are not very important, it is useful to further comment
on how our results change in the presence of such a  situation
in a real experiment. Such attractions change the picture depicted so
far as, if they are strong enough, they can cause a phase transition in
the molecule.

An effective self-attraction, like the one considered here by means of
a $6-12$ Lennard-Jones potential, arises due to hydrophobic interactions in
polypeptides and due to a suitably large concentration of
polyvalent counterions in the case of double-stranded DNA.
If this is the case, the polymer is in poor solvent conditions or
equivalently below the $\theta$ point (in the nomenclature of the models 
discussed above).
In these cases at zero stretch the force attains (in a thermodynamically long
chain) a non-zero value followed by a force plateau
for long molecules \cite{prl}. This indicates the
presence of a first order transition. We have already seen in the previous section that
in this situation (i.e. in poor solvent conditions) the two ensembles,
fixed force and fixed stretch, are not equivalent and indeed 
in the experiments the stretching curves of small DNA's or of DNA's 
in presence of a high concentration of polyvalent counterions 
in reality show peaks. A thorough explanation of this effect
includes polyelectrolyte modeling, finite size corrections
as well as dynamical effects and can be found in Refs. \cite{prl,murayama}.

A different scenario holds for ssDNA and for RNA (this scenario would
also be retraced if a self-attractive polymer such as dsDNA in presence
of condensing agents is restricted conformationally to stay in a 
quasi-two-dimensional film). In these cases the single molecule phase
transition is second order and thus at low stretch the
characteristic force curves are non-zero and then rise smoothly. Furthermore,
the attraction in ssDNA and RNA is brought forth by the base pairing interactions 
between bases far apart in the chain. If sequence disorder is neglected,
the low force regime can be written as follows:
\begin{equation} 
\zeta\sim (f-f_c)^{1/\Delta-1}
\end{equation}
where $f_c$ is the critical force. 
The exponent $\Delta$ is between $0.5$ (the value of native 
branched polymer-like
configurations arising from base stacking \cite{parisi}) 
and $1$ (for a polymer in a good solvent).
While present day experiments show that as the extension goes to 0 the force
is non-zero \cite{Tkachenko}, the data are not clean enough to allow
discriminating between the exponents above. To be noted that within a 
simplified theory \cite{Tkachenko}, it was possible to find a law
implying $\Delta=1/2$.


\end{section}

\begin{section}{Conclusions}\label{conclusions}
Here, we have revised the well known WLC model that correctly describes the behaviour under pulling of a {\it stiff} polymer. We have pointed out that its discrete version has a different large force behaviour respect to the continuous version by Marko and Siggia \cite{markosiggia}, Eq. (\ref{crossover}). We predict a {\it crossover} force between these two different regimes. It should be noticed that a recent paper by Livadaru et al. \cite{netz} reports a similar result. However, here we have given a simpler formula, Eq. (\ref{interpol}), which can be tested on real polymers as well as on numerical simulations, where self-interactions are present.

So, firstly, we have used it to fit some experimental data on dsDNA, Fig. \ref{baumann_fig}, and ssDNA, Fig. \ref{extra}. In the former case, where Eq. (\ref{markosiggia}) was already applied with success, we find that Eq. (\ref{interpol}) predicts a not trivial intra bead distance and a crossover force, which is much greater than the experimental forces. This justifies the approach by Marko and Siggia. In the latter case, we observe a clear crossover between the two regimes. In fact, now the crossover force is small and the assumption leading to Eq. (\ref{markosiggia}) is no longer justified.

Then, we have performed some Monte-Carlo simulations to verify the validity of our formula. We have analyzed different kinds of intra bead potentials. For a short range {\it repulsive} potential we have correctly found that our formula is in good agreement with the numerical data, predicting a renormalized stiffness \cite{thirumalai,podg}. For a more realistic Lennard-Jones potential the situation is more complicated. For temperatures above the $\theta$ point, Eq. (\ref{interpol}) gives a good fit. The main result is that, now, both the persistence length and the intra bead distance $b$ are renormalized by the potential. For temperatures below the $\theta$ point, our formula agrees with the numerical data only at large forces.

We have performed also simulations to examine the validity of our new formula for
the pulling of flexible polymers under quasi-fixed stretch conditions. This
kind of simulations has been done by using Molecular Dynamics methods. Like in
Monte-Carlo simulations we considered chains with repulsive and attractive
Lennard-Jones potentials but now the chains have no intrinsic stiffness. It is
found that for both types of interaction our new formula works very well at
high extension and it correctly predicts the intra bead distance. The
case of attraction is even more interesting since when the temperature is above
the $\theta$ point both the the high and low extension stretching data can be
fitted reasonably well to Eq. (\ref{interpol}). It is shown that the Marko and
Siggia formula does not fit the numerical data for the chains considered. The
reason for this is that the modest stiffness induced by the Lennard-Jones
potentials is not sufficiently large to yield the continuous WLC behaviour as
described by Eq. (\ref{markosiggia}). The results indicate that Eq.
(\ref{interpol}) can be used to characterize elastic behaviours of a much wider
range of biomolecules.
 
Finally, we have discussed how an attractive potential can modify the pulling behaviour of a polymer under poor solvent conditions.

\end{section}

\begin{section}{Acknowledgments}\label{ack}
We thank F. Seno and C. Micheletti for illuminating discussions.
\end{section}

\end{document}